# A Deployment-Oriented Framework for Explainable AI-Assisted eBPF/XDP Mitigation at the IoT Edge

Abdurrahman Tolay

Department of Computer Engineering, Istinye University, Türkiye Email:

abdurrahman.tolay@outlook.com

ORCID: 0009-0000-8855-6799

**Abstract**

Internet of Things (IoT) deployments increasingly combine heterogeneous and resource-constrained devices with weak security configurations, exposed services, limited logging, patching constraints, and long lifecycle dependencies, creating a broad attack surface across domestic, healthcare, educational, industrial, and smart-city environments. Signature-based and threshold-based controls remain useful baseline defences, but they are insufficient as standalone mechanisms in dynamic IoT networks; similarly, offline artificial intelligence (AI) benchmark performance alone does not establish operational deployability. Practical edge defence also requires efficient feature extraction, bounded inference cost, timely response, preservation of legitimate traffic, explainable decisions, and safe mitigation. This article is a conceptual framework and research agenda, and it does not report new experimental measurements. It proposes a Linux-based IoT edge-gateway architecture in which resource-aware flow-level AI-assisted detection provides a risk signal, event-level explainability supports administrator interpretation, and a bounded mitigation controller applies reversible and time-limited actions subject to critical-device safeguards. The controller updates eBPF/XDP packet-level enforcement state, records structured logs, and allows future support for validated local or privacy-aware model updates. The conceptual contribution is to connect detection, explanation, and enforcement while separating complex reasoning and policy control in user space from concise packet-handling decisions in the kernel. The framework also defines an evaluation pathway for future hardware-aware validation, including detection quality, resource cost, response timing, rollback behaviour, and legitimate-traffic preservation. Its intended value is to guide the design and assessment of safer AI-assisted IoT edge mitigation without claiming measured superiority or completed real-time performance.

| Abbreviation | Meaning |
|---|---|
| AI | Artificial Intelligence |
| XAI | Explainable Artificial Intelligence |
| IDS | Intrusion Detection System |
| IoT | Internet of Things |
| IIoT | Industrial Internet of Things |
| eBPF | programmable Linux kernel runtime historically derived from the extended Berkeley Packet Filter |
| XDP | eXpress Data Path |
| FL | Federated Learning |
| TinyML | Tiny Machine Learning |
| DDoS | Distributed Denial of Service |
| DoS | Denial of Service |
| SOC | Security Operations Center |
| SIEM | Security Information and Event Management |

Table 1: Abbreviations used in the manuscript.

# 1 Introduction

IoT systems have moved from specialized embedded deployments into everyday computing infrastructure. Smart homes use cameras, voice assistants, thermostats, locks, sensors, and appliances; healthcare environments use wearable monitors, bedside systems, remote observation devices, and connected clinical equipment; educational institutions deploy cameras, access-control systems, classroom technologies, environmental sensors, and low-cost embedded devices; industrial sites rely on IIoT devices for production monitoring, predictive maintenance, asset tracking, and process supervision; and smart-city projects depend on distributed sensing for transport, lighting, energy, environmental quality, and public safety. These deployments create operational value by improving automation, visibility, and responsiveness, but they also increase the number of endpoints that must be identified, configured, monitored, patched, and defended throughout their lifecycle (Alsaedi et al., 2020; European Union Agency for Cybersecurity, 2020; Fagan et al., 2020).

The resulting security problem is practical as much as technical. Many IoT devices ship with weak default credentials, exposed management services, limited authentication controls, restricted logging, or update mechanisms that depend on inconsistent vendor support. Low-cost devices may have limited CPU capacity, memory, storage, and power budgets, making conventional endpoint protection or deep inspection difficult to deploy. Heterogeneous protocols such as MQTT, CoAP, HTTP, proprietary cloud APIs, and vendor-specific discovery mechanisms complicate baselining and incident response. Legacy devices often remain in service after vendor maintenance has declined, and some environments cannot replace or patch devices quickly because of cost, certification, or operational continuity constraints. For these reasons, the NIST IoT cybersecurity baseline emphasizes device identification, secure configuration, logical access control, software update, data protection, cybersecurity state awareness, and documentation as core capabilities, while ENISA similarly highlights lifecycle and ecosystem risks in IoT deployments (European Union Agency for Cybersecurity, 2020; Fagan et al., 2020).

These weaknesses support a broad attack surface. IoT networks may face reconnaissance, spoofing, brute-force login attempts, unauthorized access, data leakage, malware propagation, botnet recruitment, denial-of-service attacks, distributed denial-of-service attacks, and low-rate behaviour intended to evade simple thresholds. The Mirai botnet showed how large-scale compromise can emerge from scanning and credential abuse when exposed devices remain poorly administered (Antonakakis et al., 2017; Kolias et al., 2017). More recent IoT and IIoT datasets and testbeds continue to represent diverse attack classes, including DDoS, DoS, reconnaissance, web-based attacks, brute-force attacks, spoofing, and Mirai-related traffic (Moulahi et al., 2022; Neto et al., 2023). The operational concern is therefore not only that IoT devices can be attacked, but that many deployments lack local mechanisms for observing suspicious behaviour, preserving evidence, and responding before availability, telemetry integrity, privacy, or safety are affected.

Firewalls, signatures, access-control lists, segmentation, and threshold rules remain necessary baseline controls. They can block known malicious patterns, restrict exposed services, and enforce basic network hygiene. However, they are limited as standalone defences in dynamic IoT environments where traffic changes across device models, firmware versions, user routines, cloud dependencies, and deployment contexts. Signature-based systems may miss novel or modified attacks, while threshold-based systems can be bypassed by low-rate activity or can disrupt legitimate traffic when benign behaviour changes. Centralized monitoring can add visibility, but it may introduce privacy concerns, bandwidth dependence, and delayed response when local mitigation is needed (García-Teodoro et al., 2009; Sommer & Paxson, 2010).

AI-assisted intrusion detection can improve adaptability by learning statistical patterns and behavioural deviations from traffic or telemetry. Supervised models can classify known attack types, unsupervised and semi-supervised approaches can flag abnormal behaviour, and deep learning can capture nonlinear relationships in feature spaces (Apruzzese et al., 2018; Buczak & Guven, 2016; Ferrag et al., 2020; Khraisat et al., 2019). Public datasets such as CICIoT2023, TON_IoT, Edge-IIoTset, N-BaIoT, and BoT-IoT have enabled reproducible experimentation across IoT and IIoT attack scenarios (Alsaedi et al., 2020; Koroniotis et al., 2019; Meidan et al., 2018; Moulahi et al., 2022; Neto et al., 2023; Ring et al., 2019). Yet offline predictive performance does not by itself establish that a detector is deployable on a constrained gateway. Practical deployment also requires attention to feature-extraction cost, inference latency, memory and CPU overhead, false positives, concept drift, adversarial manipulation, limited interpretability, and the incomplete connection between classification outputs and enforceable network actions (Apruzzese et al., 2018; Biggio & Roli, 2018; Gama et al., 2014; Sommer & Paxson, 2010).

A significant practical gap remains between adaptive intrusion detection and safe, low-overhead packet-level mitigation at the IoT edge. Many AI-based IDS studies concentrate on offline predictive performance, whereas many eBPF/XDP mitigation approaches emphasize efficient but comparatively static, threshold-driven, or policy-driven enforcement. This does not mean that integrated work is absent; rather, the literature remains fragmented across detection accuracy, explainability, kernel-level packet processing, and operational safety. A deployment-oriented framework must coordinate adaptive detection, explanation, policy control, reversible action, structured logging, and hardware-aware validation. It must also treat AI output as a risk signal within a controlled workflow, not as an unconstrained authority that directly blocks network traffic.

This article proposes a bounded and explainable coordination model for IoT edge defence. In the proposed architecture, a Linux-based edge gateway observes packet-header and flow-level metadata from local IoT traffic. Resource-aware flow features are extracted and passed to an AI-assisted detector that estimates risk at the event or flow level. Suspicious events are accompanied by administrator-readable explanations that connect the decision to operationally meaningful indicators, such as abnormal connection rates, unusual destination diversity, unexpected protocol use, or traffic bursts. A policy-mediated mitigation controller then selects bounded actions, such as alerting, rate limiting, temporary blocking, quarantine, or administrator escalation, according to confidence, device role, device criticality, and local policy. These actions update eBPF/XDP enforcement state so that concise packet-handling decisions can be applied efficiently in the kernel.

The separation between user space and kernel space is central to the design. Complex reasoning, feature aggregation, model inference, explanation generation, policy evaluation, and logging belong in user space, where they can be inspected, updated, and governed. The kernel path should remain concise and deterministic, using eBPF/XDP maps and actions for packet-level enforcement rather than embedding opaque model logic into per-packet processing. Mitigation actions are logged, time-limited, and reversible, with stricter safeguards for critical devices such as healthcare monitors, access-control systems, industrial controllers, or other services where false positives may create unacceptable disruption. Local retraining, TinyML extensions, or federated learning can support future adaptation, but in this manuscript they are treated as extensibility directions or optional model-update mechanisms rather than as experimentally implemented components (Banbury et al., 2021; McMahan et al., 2017; Warden & Situnayake, 2019).

This manuscript is a conceptual framework and research agenda. It does not claim completed empirical validation, measured latency, measured throughput, proven mitigation effectiveness, or demonstrated real-time performance. Its purpose is to define an architecture, design requirements, safety safeguards, and an evaluation pathway for future implementation and hardware-aware testing. The framework targets low-latency enforcement and resource-aware operation, but those targets require validation on constrained gateways under mixed benign and malicious traffic. By positioning AI-assisted detection as one input to a policy-controlled mitigation workflow, the article aims to support safer and more transparent IoT edge defence without overstating what has already been implemented.

## 1.1 Main Contributions

This article makes four contributions.

- **C1. Deployment-oriented problem formulation.** It identifies the gap between offline AI-based IDS benchmarking and operational IoT edge defence, including resource cost, mitigation readiness, explainability, concept drift, adversarial robustness, and legitimate-traffic preservation.
- **C2. Bounded edge-security architecture.** It proposes a conceptual architecture that separates a user-space control and intelligence path from an eBPF/XDP kernel enforcement path. AI produces a risk signal, and the mitigation controller applies policy-mediated, time-limited, and reversible actions.
- **C3. Safety and governance requirements.** It defines safeguards for device criticality, allowlists, confidence-aware actions, rollback, timeouts, administrator escalation, structured logs, and privacy-aware validation before model updates influence enforcement.

**C4. Hardware-aware evaluation pathway.** It specifies a future validation pathway that measures detection quality, feature-extraction cost, inference latency, memory use, CPU overhead, throughput, detection-to-enforcement latency, explanation overhead, rollback reliability, and legitimate-traffic preservation.

# 2 Background and Technical Foundations

## 2.1 IoT Security Constraints

IoT security is shaped by device heterogeneity, long operational lifecycles, fragmented ownership, and constrained hardware. A single home, school, clinic, factory, or smart-city site may contain devices from many vendors, each with different configuration interfaces, cloud dependencies, authentication mechanisms, update practices, and logging capabilities. Many deployments also include legacy or low-cost devices that remain operational after vendor support has weakened or after firmware updates become difficult to apply. These conditions complicate inventory, patching, access control, incident response, and forensic reconstruction (European Union Agency for Cybersecurity, 2020; Fagan et al., 2020).

Common weaknesses include weak default configurations, exposed management services, incomplete patching, limited endpoint logging, and restricted CPU, memory, energy, and storage resources. The NIST baseline treats securability as a device capability, including device identification, secure configuration, logical access control, software update, data protection, cybersecurity state awareness, and documentation (Fagan et al., 2020). In practice, however, many deployed IoT systems cannot fully satisfy these expectations. Gateway-level security is therefore important as a compensating control: it can observe traffic, preserve event records, segment or rate-limit risky communication, and enforce local policy. It should not be interpreted as a replacement for secure device design, strong authentication, patching, encryption, or vendor lifecycle management.

## 2.2 AI-Assisted Intrusion Detection at the IoT Edge

AI-assisted intrusion detection uses traffic or telemetry features to infer whether activity is benign, suspicious, or malicious. Supervised models learn from labelled examples of normal and attack traffic; unsupervised models identify deviations from learned normal behaviour; and semi-supervised approaches use limited labels or primarily benign data to support anomaly detection. Classical machine-learning models such as logistic regression, decision trees, random forests, support vector machines, and gradient boosting are common because they can operate on tabular flow features. Deep learning models, including autoencoders, convolutional networks, recurrent networks, and hybrid architectures, can represent more complex relationships but typically require more memory, computation, and tuning (Buczak & Guven, 2016; Ferrag et al., 2020; Khraisat et al., 2019).

For IoT edge deployment, the feature pipeline is as important as the classifier. Flow-level and packet-header-level features such as protocol, ports, flags, byte counts, packet counts, inter-arrival timing, connection rates, and destination diversity can support privacy-preserving detection while avoiding routine payload inspection. However, feature extraction itself consumes CPU cycles and memory, and delayed feature windows can reduce response timeliness. A model that performs well in an offline split may still be unsuitable for a constrained gateway if inference latency, feature-extraction overhead, memory use, false positives, or burst handling are not evaluated. Operational deployability also requires robustness to concept drift as firmware, user behaviour, and cloud services change over time, and to adversarial manipulation such as evasion, poisoning, and low-rate attacks (Apruzzese et al., 2018; Biggio & Roli, 2018; Gama et al., 2014; Sommer & Paxson, 2010).

## 2.3 Explainable AI for Operational Cybersecurity

Explainable AI is important in cybersecurity because model outputs can lead to operational consequences, including alerts, rate limits, temporary blocks, quarantine, or administrator escalation. Event-level explanations help administrators understand why a flow or device was treated as suspicious, support triage, guide rollback decisions, assist false-positive investigation, and enable later policy review. SHAP and LIME are representative methods: SHAP estimates feature contributions using a game-theoretic formulation, while LIME approximates local model behaviour around a prediction (Lundberg & Lee, 2017; Ribeiro et al., 2016). Broader XAI research emphasizes that explanations can improve transparency but must be designed for their intended users and operational context (Arrieta et al., 2020; Wang et al., 2025).

For IoT edge security, a useful explanation should translate feature attribution into administrator-readable language. For example, an explanation should indicate that a camera generated an unusual number of outbound SYN packets to many destinations, or that a thermostat communicated with unexpected external services, rather than merely reporting a high normalized feature weight. Explanation also has cost: generating feature attributions can add latency, consume resources, and complicate the response pipeline. A deployment-oriented framework must therefore decide when explanations are required, how detailed they should be, and how they support rollback, escalation, and policy refinement.

## 2.4 eBPF/XDP as an Enforcement Substrate

The programmable Linux kernel runtime now known as eBPF, historically derived from the extended Berkeley Packet Filter, allows verified programs to run safely in selected kernel contexts. XDP is an early packet-processing hook in the network receive path that can pass, drop, redirect, or otherwise account for packets before they traverse much of the networking stack. eBPF maps provide shared state between user-space controllers and kernel-resident programs, enabling controlled updates to policy state, counters, blocklists, rate-limit parameters, and monitoring data without recompiling the kernel program (Gregg, 2019; Høiland-Jørgensen et al., 2018; Vieira et al., 2020).

For an IoT edge gateway, eBPF/XDP is attractive because concise packet-handling logic can enforce decisions close to the network interface. Typical actions include packet pass, packet drop, redirect, traffic accounting, monitoring, and bounded policy-state updates. The safety and maintainability of this substrate depend on keeping kernel-resident logic concise and verifiable. Complex reasoning, feature aggregation, model inference, explanation generation, policy evaluation, and governance should remain in user space, while the kernel path should apply compact enforcement decisions through maps and well-defined XDP actions. This distinction between user-space control and intelligence and kernel-level enforcement is central to the framework proposed in this manuscript.

## 2.5 TinyML and Federated Learning as Extensibility Directions

TinyML and federated learning are relevant supporting directions, but they are not implemented components in the present conceptual manuscript. TinyML focuses on compact models, quantization, pruning, and efficient inference for highly constrained devices, and it may support device-level anomaly scoring or compact local classification in future extensions (Banbury et al., 2021; Warden & Situnayake, 2019). Its limitations include reduced model capacity, restricted feature availability, energy constraints, and possible loss of attack-class granularity.

Federated learning offers a privacy-aware update strategy in which participating sites exchange model updates rather than centralizing raw traffic data (McMahan et al., 2017). For IoT intrusion detection, it may help distributed gateways adapt to local traffic while reducing raw-data sharing, but it introduces non-IID data, communication overhead, unreliable clients, poisoning risk, aggregation challenges, and model-governance requirements (Biggio & Roli, 2018; Khraisat et al., 2024; Zhang et al., 2025). Secure aggregation, client validation, local acceptance testing, and model rollback are therefore necessary before federated updates influence enforcement. In this manuscript, TinyML and federated learning are treated as extensibility directions that can support future adaptation, not as experimentally validated parts of the proposed architecture.

# 3 Related Work

## 3.1 IoT Intrusion-Detection Datasets and Evaluation Limitations

Public datasets have supported reproducible IDS research, but they differ substantially in device composition, traffic generation, attack coverage, labels, and feature definitions. CICIoT2023 provides large-scale IoT attack traffic across many classes, including DDoS, DoS, reconnaissance, web-based attacks, brute-force attacks, spoofing, and Mirai-related traffic (Neto et al., 2023). TON_IoT includes telemetry and network data for IoT and IIoT contexts (Alsaedi et al., 2020). Edge-IIoTset was designed for edge-IIoT cybersecurity evaluation and includes multiple attack categories and heterogeneous data sources (Moulahi et al., 2022). N-BaIoT focuses on network-based detection of IoT botnet attacks using device-specific traffic traces (Meidan et al., 2018). BoT-IoT provides botnet-oriented traffic for IoT network security evaluation (Koroniotis et al., 2019). UNSW-NB15 is often used as a supplementary benchmark, but it is not IoT-specific and should not be treated as a substitute for IoT deployment validation (Moustafa & Slay, 2015).

Dataset use requires caution. Label mismatch, feature-definition mismatch, leakage risk, class imbalance, synthetic artefacts, and dataset-specific traffic patterns can inflate reported performance or reduce transferability to live networks (Ring et al., 2019; Sommer & Paxson, 2010). Cross-dataset evaluation can expose some of these weaknesses, but it cannot fully replace hardware-in-the-loop validation on constrained gateways with mixed benign and malicious traffic. For the present manuscript, datasets are useful for model development and initial comparison, while deployment claims require measurements of feature-extraction cost, inference latency, CPU use, memory use, throughput, explanation overhead, detection-to-enforcement latency, rollback reliability, and legitimate-traffic preservation.

## 3.2 AI-Based IoT IDS Research

AI-based IoT IDS research includes classical machine learning, deep learning, lightweight models, and anomaly-detection approaches. Classical models are often attractive for edge settings because they can operate on flow-level tabular features and may be easier to interpret or compress. Deep learning can improve representation capacity and has been widely studied for cybersecurity intrusion detection, but it can increase computational cost and may require larger datasets and careful tuning (Ferrag et al., 2020; Khraisat et al., 2019). Anomaly-detection approaches are valuable when labelled attacks are incomplete, although they can produce false positives when benign device behaviour changes.

A recurring limitation is the overemphasis on predictive scores such as accuracy, precision, recall, or F1-score without corresponding deployment metrics. Offline evaluation often omits feature-extraction overhead, inference latency, memory use, CPU overhead, explanation overhead, mitigation readiness, and behaviour under traffic bursts. It may also understate concept drift and adversarial manipulation risks (Apruzzese et al., 2018; Biggio & Roli, 2018; Gama et al., 2014; Sommer & Paxson, 2010). For an IoT edge gateway, the central question is not only whether a classifier can distinguish labelled attack traffic, but whether it can support safe, timely, explainable, and reversible operational decisions.

## 3.3 eBPF/XDP-Based Detection and Enforcement

Bachl, Fabini, and Zseby demonstrate a flow-based machine-learning IDS implemented entirely in eBPF using a decision tree (Bachl et al., 2021). Their work is highly relevant because it shows that some forms of ML inference can be placed inside eBPF. It should not, however, be misread as an IoT-specific framework or as a complete governance model for bounded mitigation, explanation, rollback, or critical-device handling.

Feraudo et al. apply Manufacturer Usage Description (MUD) concepts and eBPF-based traffic filtering to mitigate IoT botnet DDoS attacks (Feraudo et al., 2024). Their work demonstrates IoT-specific policy enforcement and rate-control potential, particularly for constraining device traffic according to expected communication behaviour. It does not explicitly provide the same XAI-driven bounded controller, rollback semantics, or privacy-aware model-update validation proposed in the present manuscript.

Tolay reports eBPF-based IoT DDoS mitigation using rate-based detection and evaluates in-kernel mitigation with Docker-based simulation and Raspberry Pi hardware (Tolay, 2025). That work is relevant because it addresses IoT edge mitigation and hardware-aware evaluation. The present conceptual manuscript seeks to extend this direction toward AI-assisted risk assessment, event-level explainability, bounded controller logic, critical-device safeguards, and broader evaluation requirements. This distinction is conceptual and architectural; no measured superiority is claimed here.

Vieira et al. provide a technical foundation for eBPF/XDP concepts, fast packet processing,

kernel programmability, maps, hooks, challenges, and applications (Vieira et al., 2020). Together with XDP and BPF systems work, this literature supports the design choice of keeping packet enforcement concise in the kernel while leaving complex intelligence and policy management in user space (Gregg, 2019; Høiland-Jørgensen et al., 2018).

## 3.4 Integrated AI and eBPF/XDP Frameworks

Integrated AI and eBPF/XDP work already exists, and this prevents the present manuscript from claiming that AI-assisted eBPF/XDP integration is absent from the literature. Farasat, Kim, and Posegga propose SmartX Intelligent Sec, a framework combining eBPF/XDP-based packet capture and malicious-traffic filtering with a BiLSTM classifier for network-threat detection (Farasat et al., 2024). SmartX focuses on automated threat detection and filtering for broader ICT infrastructures. The present manuscript differs by focusing on an IoT-edge conceptual architecture with explicit bounded policy mediation, reversible actions, timeouts, critical-device safeguards, administrator escalation, structured logging, drift-aware validation, privacy-aware model-update safeguards, legitimate-traffic preservation, and a deployment-oriented evaluation pathway. Because the present manuscript is conceptual, it does not imply measured superiority over SmartX.

McAuley et al. describe a human-centred home network security architecture that includes eBPF-IoT components for IoT monitoring and control (McAuley et al., 2022). Their architecture includes eBPF-Mon for traffic statistics and ML-feature extraction, eBPF-IoT-MUD for traffic-policy enforcement, and XDP-based packet dropping and traffic management. This is important related work because it connects IoT monitoring, feature extraction, and policy enforcement. The present manuscript differentiates itself through an explicit explainability layer, a bounded mitigation controller, critical-device policies, rollback semantics, drift-aware update validation, and a formal evaluation pathway for future hardware-aware validation.

## 3.5 XAI, TinyML, and Federated Learning in IoT IDS

XAI, TinyML, and federated learning form complementary research streams for IoT IDS. XAI improves transparency and can support triage, policy review, and false-positive investigation, but it introduces explanation overhead and usability concerns when feature attributions are not translated into operational language (Arrieta et al., 2020; Lundberg & Lee, 2017; Ribeiro et al., 2016; Wang et al., 2025). TinyML supports resource-aware inference on constrained devices, but compact models may reduce model capacity, feature richness, or attack granularity (Banbury et al., 2021; Warden & Situnayake, 2019). Federated learning supports privacy-aware collaboration across distributed sites, but it introduces non-IID data, communication overhead, poisoning risk, aggregation challenges, and model-governance requirements (Biggio & Roli, 2018; Khraisat et al., 2024; McMahan et al., 2017; Zhang et al., 2025). These streams are relevant to deployable IoT security, but they do not independently solve the detection-to-enforcement gap.

## 3.6 Comparative Analysis of Closely Related Work

Table 2: Comparison of Closely Related IoT Edge-Security and eBPF/XDP Approaches

| Study | Context | ML | XAI | eBPF/XDP role | Act. | Roll. | Crit. | Drift | HW | Main limitation |
|---|---|---|---|---|---|---|---|---|---|---|
| Bachl et al. (2021) | Flow-based IDS | Y | N | In-kernel decision-tree IDS | NS | NS | NS | NS | NS | Not IoT-specific; no explainable bounded-mitigation governance. |
| McAuley et al. (2022) | Home IoT security | P | NS | Monitoring, feature extraction, MUD enforcement, and XDP control | P | NS | NS | NS | P | No explicit XAI, rollback semantics, critical-device safeguards, or drift-aware update validation. |
| Feraudo et al. (2024) | IoT botnet DDoS mitigation | N | N | MUD and eBPF filtering | P | NS | NS | NS | P | IoT-specific filtering without AI-assisted explainable controller governance. |
| Farasat et al. (2024) | ICT threat detection | Y | N | eBPF/XDP capture and filtering | P | NS | NS | NS | NS | ML plus eBPF/XDP filtering without IoT-criticality, rollback, or model-governance focus. |
| Tolay (2025) | IoT edge DDoS mitigation | N | N | Rate-based eBPF mitigation | P | P | NS | N | Y | Rate-based mitigation without AI-assisted risk scoring, XAI, or governed model lifecycle. |
| Proposed framework | IoT edge defence | SC | SC | Two-plane monitoring and enforcement architecture | SC | SC | SC | SC | Planned | Conceptual framework; implementation and empirical validation remain future work. |

*Notes: Y = yes; N = no; P = partial; SC = specified conceptually; NS = not specified; Planned = future validation planned.*

The comparison indicates that the proposed contribution is not the first integration of AI and eBPF/XDP, but a more explicit governance model for bounded, explainable, reversible, and hardware-aware IoT-edge mitigation.

The comparison shows that AI-assisted eBPF/XDP integration already exists, eBPF-based IoT monitoring and policy enforcement already exist, and MUD-based IoT mitigation already exists (Farasat et al., 2024; Feraudo et al., 2024; McAuley et al., 2022). The defensible contribution of the present framework is therefore not technology stacking or first integration. It is the formal coordination of event-level explanation, bounded mitigation, rollback, critical-device safeguards, administrator escalation, structured logging, privacy-aware model-update validation, legitimate-traffic preservation, and hardware-aware evaluation for constrained IoT-edge environments.

# 4 Research Gap

Existing work demonstrates important progress in AI-based intrusion detection, eBPF-based flow analysis, XDP enforcement, IoT-specific MUD filtering, and integrated ML-assisted packet filtering (Bachl et al., 2021; Farasat et al., 2024; Feraudo et al., 2024; McAuley et al., 2022; Tolay, 2025). This literature establishes that adaptive detection and kernel-level packet control are both technically feasible, and that they can be combined in several ways. The remaining gap is therefore not the absence of AI or eBPF/XDP integration.

The literature remains fragmented across detection, explanation, safe automation, reversible enforcement, device criticality, concept drift, model-update governance, hardware-aware evaluation, and legitimate-traffic preservation. AI-based IDS studies often provide strong predictive evaluation but do not fully specify how classifications become bounded and reversible mitigation actions. eBPF/XDP and MUD-based systems provide efficient enforcement or policy control, but they do not always incorporate event-level explanations, drift-aware model validation, privacy-aware update governance, or critical-device safeguards. Integrated ML-assisted eBPF/XDP systems demonstrate important architectural progress, but their stated scope does not fully cover the IoT-edge governance requirements emphasized in this manuscript.

The research gap addressed here is the need for a sufficiently explicit, bounded, explainable, and deployment-oriented coordination model for constrained IoT-edge environments. Such a model should treat AI outputs as risk signals rather than direct blocking commands; mediate actions through policy; support timeouts, rollback, and escalation; preserve legitimate traffic where possible; account for device criticality; record structured evidence; and define how future implementations should be validated on hardware. This gap is practical and architectural rather than purely algorithmic.

### 4.1 Originality of the Proposed Framework

The originality of the proposed framework is architectural and operational rather than algorithmic. It does not claim to introduce a new AI classifier, a new XDP primitive, or the first AI-eBPF/XDP integration. Its contribution is the formal coordination of flow-level AI-assisted risk assessment, event-level explanation, policy-mediated mitigation, action timeouts, rollback, device-criticality handling, administrator escalation, structured logging, privacy-aware update validation, drift-aware safeguards, legitimate-traffic preservation, and hardware-aware evaluation.

AI is positioned as a risk signal inside a governed mitigation pipeline. The user-space path performs feature aggregation, inference, explanation, policy evaluation, logging, and update validation. The eBPF/XDP path applies concise packet-level enforcement state through maps and bounded actions. This separation allows the framework to benefit from adaptive detection while preserving operational control over mitigation decisions. The contribution is therefore a cautious coordination model for future implementation and validation, not a claim of measured superiority over existing AI-assisted eBPF/XDP systems.

## 5 Threat Model and System Assumptions

### 5.1 Protected Assets and Security Objectives

The protected assets are IoT-device availability, integrity of permitted communication, local-network availability, gateway forwarding capacity, device communication profiles, telemetry integrity, policy state stored in eBPF maps, structured event logs, model versions and update metadata, administrator visibility, and continuity of critical-device communication. These assets matter because an IoT edge-security mechanism can fail not only by missing an attack, but also by disrupting legitimate services, losing evidence, corrupting enforcement state, or allowing an unvalidated model update to influence mitigation.

The main security objectives are to detect suspicious flow-level behaviour, apply bounded and reversible mitigation, preserve legitimate traffic wherever possible, avoid unsafe disruption of critical devices, preserve evidence for review, prevent unvalidated model updates from influencing enforcement, and limit privacy exposure by avoiding routine payload inspection.

The framework is a compensating network-level control. It does not replace secure firmware, patching, strong authentication, encryption, segmentation, vendor lifecycle management, endpoint hardening, or incident-response procedures.

## 5.2 Deployment Boundary and Traffic Visibility

The conceptual framework assumes a Linux-based IoT edge gateway with eBPF/XDP support, IoT traffic traversing an observable enforcement point, packet-header-level and flow-level telemetry collection, user-space control and intelligence components running on or near the gateway, current policy state stored in one or more eBPF maps, and structured logs stored locally or exported to an approved monitoring workflow. The framework's enforcement capability depends directly on traffic visibility and on whether the gateway is placed on the relevant forwarding path.

Traffic that bypasses the gateway cannot be observed or mitigated by the framework. East-west traffic within the same local segment may not traverse the enforcement point unless the network topology, VLAN design, bridge configuration, switch architecture, or wireless-controller path directs it through an observable interface. Encrypted payloads do not prevent header-level and flow-level analysis, but payload-dependent attacks may not be detectable without additional inspection. NAT, IPv6, VLANs, wireless interfaces, multiple network interfaces, and driver support may affect implementation details. XDP redirect support and native-driver support can also vary by deployment. Identity based only on source IP or MAC address can be weakened by spoofing; future implementations should combine network identifiers with inventory, interface, segment, and policy context where possible.

## 5.3 Trusted Components and Trust Boundaries

The primary trusted components are the Linux kernel and eBPF verifier, the loaded eBPF/XDP program, the authorized eBPF map update interface, the user-space policy controller, the rollback manager, approved policy templates, the administrator authentication mechanism, the local model-validation process, and the logging integrity mechanism. The framework assumes that these components are not already fully compromised.

A fully compromised gateway, malicious kernel module, compromised administrator account, unauthorized root access, or attacker-controlled policy controller can undermine enforcement and is outside the primary protection scope. Future implementations should still reduce these risks through least privilege, signed or approved program deployment, restricted map-update permissions, separation of duties, protected logs, authenticated administrative access, version control, rollback, and auditable policy changes. These controls are design requirements for future implementation, not completed safeguards reported by this article.

### 5.4 Attacker Capabilities

The attacker may be an external network attacker, an attacker controlling one or more compromised IoT devices, an attacker attempting to recruit devices into a botnet, an attacker attempting lateral movement, or an attacker generating high-rate or low-rate traffic. The attacker may adapt to static thresholds, mimic benign behaviour, attempt source-address spoofing where network conditions allow, or try to exhaust gateway CPU, memory, bandwidth, telemetry buffers, eBPF-map capacity, logging capacity, explanation workload, or administrator attention. In future federated-learning scenarios, an attacker may attempt poisoning or malicious model-update influence. The attacker may also exploit benign concept drift or behavioural changes to trigger false positives. The framework does not assume that attack traffic is always high-volume.

### 5.5 Addressed Attack Classes

The framework is intended to support detection and bounded response for scanning and reconnaissance, abnormal connection bursts, repeated failed connections, brute-force attempts where observable through network behaviour, DoS and DDoS traffic, botnet-like outbound communication, unusual destination diversity, unexpected protocol or port use, rate anomalies, selected spoofing indicators, low-rate evasive behaviour, compromised-device communication patterns, and some forms of lateral movement where traffic crosses the enforcement point. Support varies according to available flow features, visibility, traffic-window design, model quality, network topology, policy configuration, encryption, and implementation choices. The framework therefore supports bounded response to observable behavioural indicators; it does not promise comprehensive detection.

### 5.6 Out-of-Scope and Partially Addressed Conditions

The framework does not claim to prevent or fully detect physical tampering, hardware implants, radio-frequency jamming, power disruption, attacks that bypass the gateway, attacks entirely contained within an unobserved local segment, fully compromised gateway kernels, malicious root administrators, vulnerabilities inside device firmware that do not create observable network behaviour, application-layer attacks requiring payload semantics when payload inspection is disabled, encrypted-content attacks that cannot be inferred from metadata, zero-day attacks that do not produce detectable behavioural deviations, universal attribution of spoofed identities, guaranteed protection against all adversarial-ML attacks, or guaranteed safety of federated learning without additional validation controls.

The framework can complement, but not replace, segmentation, network access control, Manufacturer Usage Description policies, firewall rules, secure boot, endpoint hardening, certificate-based authentication, patch management, firmware validation, physical security, and incident-response procedures. Conditions outside the observable network path or outside the trusted gateway boundary remain only partially addressed or out of scope.

## 5.7 Critical-Device Policy Model

Critical devices are assets whose disruption may affect safety, clinical workflows, physical access, industrial operation, or emergency communication. Examples include healthcare monitors, infusion-related systems, access-control devices, industrial controllers, safety-related sensors, and emergency communication systems. Future deployments should assign a policy class to each protected asset, such as standard, sensitive, critical, or safety-critical.

Criticality affects action thresholds, permitted responses, timeout duration, escalation requirements, rollback urgency, logging detail, and administrator-confirmation requirements. The most disruptive actions should normally require higher confidence, narrower scope, or human confirmation for critical and safety-critical devices. Exact rules depend on deployment risk analysis, regulatory constraints, organizational policy, and the availability requirements of the protected environment.

## 5.8 Adversarial-ML and Resource-Exhaustion Considerations

The threat model includes evasion during inference, poisoning during training or model update, benign concept drift, malicious manipulation of flow features, telemetry flooding, alert flooding, explanation-workload exhaustion, map-capacity exhaustion, logging exhaustion, and administrator-attention exhaustion. These risks are important because an attacker may target the monitoring and decision pipeline rather than only the protected IoT devices.

The intended conceptual mitigations include bounded telemetry sampling, flow aggregation, map-capacity controls, expiration of temporary state, rate-limited alerting, explanation generation only for selected events, suspicious-update validation, model versioning, local acceptance testing, rollback, conservative fallback policy, and administrator review. Future implementation should validate whether these measures reduce overload and manipulation risk under realistic traffic conditions. This article does not claim validated protection against adversarial ML or resource-exhaustion attacks.

### 5.9 Privacy Assumptions and Data Minimization

The default design prioritizes packet-header-level and flow-level metadata rather than payload inspection. This supports data minimization by limiting routine collection to metadata needed for traffic analysis, policy decisions, and auditability. Local processing can reduce unnecessary transfer of sensitive traffic information, while structured logging can record device identifiers, flow metadata, risk scores, policy decisions, action state, and explanation summaries without storing raw payloads by default.

Privacy controls in future deployments should consider retention limits, access control, purpose limitation, protected logs, optional export to SIEM workflows, and future federated-learning updates without raw-data centralization. Privacy requirements depend on jurisdiction, organizational policy, deployment context, and the sensitivity of device traffic. The framework provides a technical data-minimization orientation, not a legal compliance claim.

Table 3: Threat Model Summary

| Element | Assumption or capability | Security relevance | Framework response | Limitation |
|---|---|---|---|---|
| External attacker | Sends scans, floods, spoofed traffic, or access attempts. | Can affect availability and expose weak services. | Flow monitoring, risk scoring, bounded enforcement, logging. | Only traffic crossing the enforcement point is visible. |
| Compromised IoT device | Generates abnormal outbound or lateral traffic. | May recruit botnets or attack local peers. | Device-aware policy, temporary limits, alerts, rollback. | Firmware compromise itself may remain unresolved. |
| Low-rate adaptive attacker | Attempts to evade static thresholds. | Can avoid simple rate rules. | Flow-window analysis and drift-aware evaluation. | Detection depends on feature quality and model validation. |
| Source spoofing | Uses false source identifiers where allowed. | Weakens IP- or MAC-only attribution. | Combine identifiers with segment, interface, inventory, and policy context. | Universal attribution is not claimed. |
| Botnet-like outbound traffic | Contacts unusual destinations or produces command-and-control-like patterns. | Indicates possible device compromise. | Destination-diversity features, alerts, bounded response. | Encrypted payload semantics may remain hidden. |
| Gateway resource exhaustion | Targets CPU, memory, bandwidth, maps, buffers, or logs. | Can degrade forwarding and monitoring. | Bounded telemetry, map-capacity controls, rate-limited alerts. | Future stress testing is required. |

Table 3: Threat Model Summary (continued)

| Element | Assumption or capability | Security relevance | Framework response | Limitation |
|---|---|---|---|---|
| Telemetry or alert flooding | Produces excessive events. | Can hide incidents or exhaust administrators. | Aggregation, sampling, rate-limited alerting, structured logs. | May reduce visibility during overload. |
| Model-update poisoning | Attempts to influence future model updates. | Can corrupt enforcement decisions. | Versioning, local acceptance tests, suspicious-update validation. | Federated-learning safety is not guaranteed. |
| Concept drift | Benign behaviour changes over time. | Can raise false positives or reduce detection. | Drift alerts, conservative fallback, retraining review. | Requires deployment-specific validation. |
| Critical-device false positive | Mitigation affects safety-sensitive service. | Can disrupt clinical, industrial, or access-control workflows. | Criticality classes, narrower actions, escalation, rollback. | Exact policy depends on local risk analysis. |
| Gateway compromise | Kernel, root account, or controller is compromised. | Undermines enforcement and evidence. | Future hardening should use least privilege and protected logs. | Fully compromised gateway is outside primary scope. |
| Traffic bypassing enforcement point | Traffic avoids the gateway. | Prevents observation and mitigation. | Deployment planning must place gateway on relevant paths. | Bypassed traffic is out of scope. |

Table 3: Threat Model Summary (continued)

| Element | Assumption or capability | Security relevance | Framework response | Limitation |
|---|---|---|---|---|
| Encrypted traffic | Payload contents are unavailable by default. | Limits payload-dependent attack detection. | Header and flow metadata analysis. | Encrypted-content attacks may not be inferred. |
| Unobserved east-west traffic | Local traffic does not traverse the gateway interface. | Lateral movement may be invisible. | VLAN, bridge, switch, or controller design can improve visibility. | Topology-dependent and partially addressed. |

Table 4: Threat-to-Control Mapping for the Proposed Conceptual Framework

| **Threat or failure mode** | **Observable indicators** | **Primary framework component** | **Candidate bounded response** | **Residual risk** | **Future validation requirement** |
|---|---|---|---|---|---|
| Scanning | Many connection attempts or destination ports. | Flow aggregator and risk scorer. | Monitor, alert, temporary rate limit, or port restriction. | Slow scans may blend with benign traffic. | Validate across scan rates and device baselines. |
| Brute-force behaviour | Repeated failed connections where observable. | Feature extractor and controller. | Alert, temporary limit, or destination-port restriction. | Application failures may be encrypted or unavailable. | Validate with protocol-specific observability constraints. |
| UDP flood | High UDP packet or byte rate. | XDP data path and counters. | Candidate response: accounting, temporary drop, or rate limit. | Source spoofing and collateral damage remain possible. | Measure throughput, false blocking, and rollback. |
| TCP SYN flood | High SYN rate or incomplete handshakes. | XDP counters and policy maps. | Candidate response: temporary drop or rate control. | Legitimate bursts may be affected. | Validate under mixed benign and attack traffic. |
| Low-rate DoS | Sustained abnormal load below static thresholds. | Risk scorer and drift-aware evaluation. | Monitor, alert, narrow rate limit where policy permits. | May be hard to distinguish from usage changes. | Test low-rate variants and calibration. |
| Botnet-like outbound traffic | Unusual destinations, ports, or bursts. | Risk scorer and controller. | Alert, restrict selected destinations, or request quarantine. | Payload encryption may hide command semantics. | Validate destination-diversity and baseline features. |

Table 4: Threat-to-Control Mapping for the Proposed Conceptual Framework (continued)

| **Threat or failure mode** | **Observable indicators** | **Primary framework component** | **Candidate bounded response** | **Residual risk** | **Future validation requirement** |
|---|---|---|---|---|---|
| Unusual destination diversity | Many unfamiliar external endpoints. | Feature extractor. | Monitor, alert, temporary restriction. | Cloud services may change legitimately. | Separate benign drift from compromise. |
| Spoofing attempt | Inconsistent identifiers or impossible segment context. | Policy controller and inventory context. | Alert, stricter monitoring, or narrow restriction. | Attribution may remain uncertain. | Validate with IP, MAC, VLAN, and interface context. |
| Lateral movement across observable interfaces | Internal scanning or unexpected peer communication. | Gateway monitoring and controller. | Alert, segment-aware restriction, or escalation. | Unobserved east-west paths remain invisible. | Test bridge, VLAN, and wireless-controller scenarios. |
| Telemetry flooding | Excessive telemetry events or buffer pressure. | Telemetry pipeline. | Sampling, aggregation, queue limits, rate-limited alerts. | Reduced detail during overload. | Stress-test buffers, drops, and recovery. |
| Alert flooding | Many alerts overwhelm review. | Logging and alerting workflow. | Deduplication, prioritization, rate-limited alerting. | Important events may be delayed. | Measure alert rate, suppression, and audit completeness. |
| Map-capacity exhaustion | High occupancy in eBPF map. | eBPF map and controller. | Expiration, bounded entries, conservative fallback. | Policy state may be incomplete under stress. | Validate map occupancy and failure behaviour. |

Table 4: Threat-to-Control Mapping for the Proposed Conceptual Framework (continued)

| **Threat or failure mode** | **Observable indicators** | **Primary framework component** | **Candidate bounded response** | **Residual risk** | **Future validation requirement** |
|---|---|---|---|---|---|
| Explanation-workload exhaustion | Too many events requiring XAI. | XAI module and controller. | Explain selected events; sample benign flows. | Some events may lack immediate explanation. | Measure explanation queueing and latency separately. |
| Poisoned model update | Suspicious update or degraded local test result. | Model-validation process. | Reject update, retain previous model, rollback. | Sophisticated poisoning may evade checks. | Validate acceptance tests and rollback. |
| Benign concept drift | Firmware, cloud, or usage changes. | Drift monitoring and policy controller. | Conservative fallback, retraining review, administrator alert. | False positives may still occur. | Evaluate temporal windows and device-separated splits. |
| False positive affecting a critical device | Mitigation targets protected asset. | Critical-device policy model. | Escalate, narrow action, shorter timeout, or rollback. | Safety risk cannot be eliminated. | Validate policy classes, override, and recovery. |

# 6 Design Requirements and Future Validation Criteria

A deployable framework must satisfy requirements that go beyond predictive accuracy. Because the present article is conceptual, the requirements in Table 5 are expressed as future validation obligations rather than experimentally demonstrated thresholds.

Table 5: Design Requirements and Future Validation Obligations

| ID | Requirement | Rationale | Future measurement | Reporting expectation |
|---|---|---|---|---|
| R1 | Low telemetry-aggregation latency | Flow evidence must be available quickly enough to support timely analysis. | Median, p95, and p99 telemetry-aggregation latency under benign, burst, and attack conditions. | Report window design, traffic load, and collection mechanism. |
| R2 | Low feature-extraction cost | Feature computation must not overload the gateway. | CPU cost, memory overhead, and processing latency per flow window. | Report exact feature list and live availability. |
| R3 | Bounded inference latency | AI-assisted risk scoring must complete within a useful operational interval. | Median, p95, and p99 inference latency under varying load. | Report model type, model size, runtime, hardware, and batch assumptions. |
| R4 | Low memory footprint | The system must fit on Raspberry Pi-class or comparable gateways. | Peak and steady-state RAM use; model size; map memory; telemetry-buffer memory. | Separate kernel and user-space memory where feasible. |
| R5 | Detection quality with low false-positive risk | Detection must preserve legitimate services. | Macro-F1, weighted F1, per-class recall, false-positive rate, false-negative rate, confusion matrix, calibration quality. | Report random, time-separated, device-separated, and cross-dataset results where feasible. |

Table 5: Design Requirements and Future Validation Obligations (continued)

| ID | Requirement | Rationale | Future measurement | Reporting expectation |
|---|---|---|---|---|
| R6 | Packet-processing overhead | The XDP fast path must not impose excessive forwarding cost. | Throughput, packet rate, CPU use, and packet loss for pass-through, accounting, lookup, drop, redirect, and rate-control cases where implemented. | Compare against baseline forwarding. |
| R7 | Detection-to-enforcement latency | The controller must update bounded enforcement state promptly. | Median, p95, and p99 interval from satisfied mitigation condition to successful map or external-control update. | Measure separately from explanation latency. |
| R8 | Explanation quality and overhead | Explanations should support review without blocking urgent bounded response. | Explanation latency, stability, feature relevance, faithfulness where feasible, and expert usefulness. | State whether explanation is asynchronous and identify the model-specific method. |
| R9 | Reversible mitigation | Temporary actions must expire, narrow, or roll back correctly. | Rollback success rate, timeout accuracy, state-verification result, and recovery time. | Report action type and affected scope. |

Table 5: Design Requirements and Future Validation Obligations (continued)

| ID | Requirement | Rationale | Future measurement | Reporting expectation |
|---|---|---|---|---|
| R10 | Critical-device preservation | Mitigation must not disrupt safety-sensitive services unnecessarily. | False-blocking rate for protected devices; escalation behaviour; policy-exception handling. | Report policy class and permitted action set. |
| R11 | Robustness to concept drift | Behaviour changes must not cause unsafe enforcement. | Performance across temporal windows, drift alerts, retraining triggers, fallback behaviour. | Separate benign drift from attack behaviour. |
| R12 | Robustness to adversarial manipulation | Adaptive attackers may evade thresholds or poison updates. | Low-rate attacks, feature manipulation, evasion tests, poisoning-aware checks, update rejection, and fallback behaviour. | Separate inference-time evasion from training-time poisoning. |
| R13 | Resource-exhaustion resilience | Attackers may overload telemetry, maps, logs, explanations, or alerting. | Map occupancy, buffer drops, queue pressure, log volume, explanation workload, alert rate, CPU and memory under stress. | Report bounded-capacity controls and failure behaviour. |

Table 5: Design Requirements and Future Validation Obligations (continued)

| ID | Requirement | Rationale | Future measurement | Reporting expectation |
|---|---|---|---|---|
| R14 | Privacy-aware model-update validation | Updates must not centralize raw traffic unnecessarily or affect enforcement without checks. | Update validation outcome, model version, acceptance test, rollback result, communication overhead where FL is used. | Distinguish local retraining from federated update experiments. |
| R15 | Security-workflow compatibility | The framework should support administrator review and incident response. | Structured-log completeness, alert content, export format, override handling, and auditability. | Report integration boundaries without claiming full SIEM implementation unless tested. |

These requirements are interdependent. Lower latency must not come at the expense of safety, aggressive mitigation must not undermine legitimate traffic, richer explanations must not block urgent bounded action, and model updates must not bypass policy validation. Future experiments should therefore report trade-offs rather than only best-case detection or throughput metrics.

# 7 Proposed Framework

The proposed framework is a conceptual IoT edge-security architecture that separates packet-level enforcement from user-space intelligence. It is designed around two coordinated paths rather than a single sequential pipeline. The kernel-resident data path handles packets early and applies the current bounded policy state. The user-space control and intelligence path aggregates telemetry, evaluates risk, selects policy-mediated actions, updates eBPF maps, records evidence, and supports asynchronous explanation and rollback.

Incoming packets reach the XDP hook early in the receive path. The XDP program consults current kernel-resident policy state stored in one or more eBPF maps and applies concise actions such as pass, drop, redirect, accounting, or implementation-specific bounded traffic-control logic. The AI model does not synchronously classify every packet before the XDP hook. Instead, telemetry and flow summaries are exported or aggregated for user-space analysis. The user-space controller evaluates risk and updates eBPF maps so that subsequent packets are handled according to revised policy state. The framework does not assume that the first malicious packet is blocked. It targets rapid flow-level detection and bounded policy updates so that subsequent packets associated with suspicious behaviour can be mitigated efficiently in the kernel.

The kernel-resident data path is therefore: incoming packet → XDP hook → consult eBPF maps → apply current policy state → pass, drop, redirect, account, or apply bounded traffic-control logic. The user-space control and intelligence path is: telemetry export or counters → flow aggregation → resource-aware feature extraction → AI-assisted risk scoring → policy-mediated mitigation controller → eBPF map update → structured logging → optional asynchronous explanation → timeout, rollback, or escalation. These paths interact through eBPF maps, counters, flow summaries, and controlled policy updates.

## 7.1 Architectural Overview

Figure 1 presents the overall two-plane architecture separating the kernel-resident data path from the user-space control and intelligence plane.

The framework separates responsibilities between a kernel-resident data plane and a user-space control and intelligence plane. The kernel-resident data plane processes packets early, remains concise, consults eBPF maps, applies deterministic enforcement state, maintains counters, and exposes telemetry where appropriate. It avoids complex model logic by default unless a future implementation explicitly validates a simple in-kernel model for safety, performance, and maintainability.

The user-space control and intelligence plane aggregates flow evidence, extracts features, calculates AI-assisted risk scores, evaluates device role and criticality, applies policy logic, updates eBPF maps, manages timeouts and rollback, generates explanations, produces logs and alerts, and validates future model updates. The conceptual framework intentionally keeps complex reasoning outside the kernel for governability, maintainability, and safety. Earlier work demonstrates that certain simple models can run inside eBPF, including a decision-tree-based flow IDS (Bachl et al., 2021). The present framework nevertheless uses user-space inference as its default architecture because it prioritises explainability, policy governance, rollback, model-update validation, and operational maintainability.

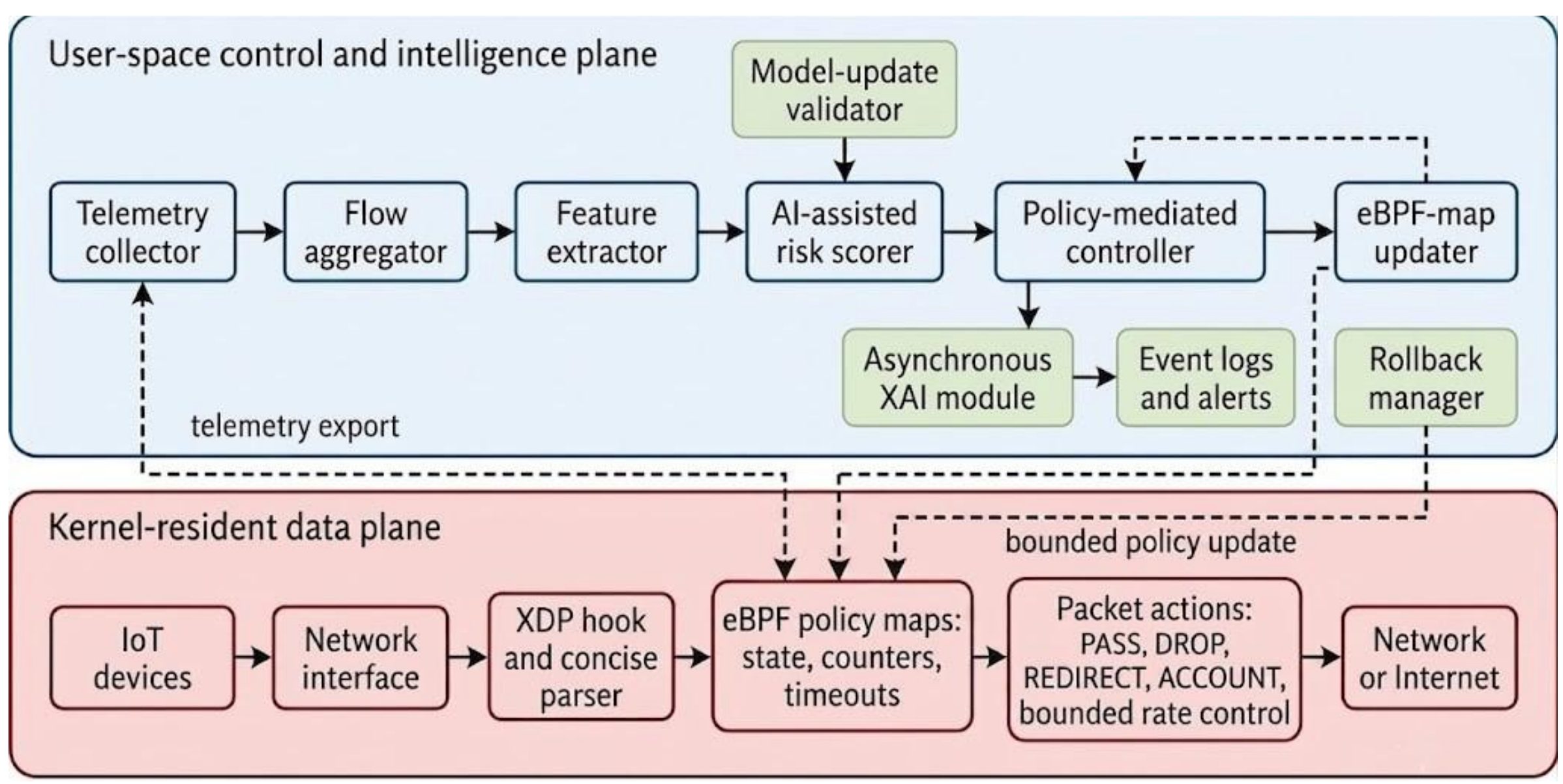

Figure 1: Two-plane architecture separating the kernel-resident XDP data path from the user-space control and intelligence plane. Packet enforcement remains on the fast path, while risk scoring, policy mediation, explanation, rollback, and validated model updates are managed in user space.

## 7.2 Kernel-Resident XDP Data Path

The kernel-resident XDP data path begins when a packet arrives at the network interface. The XDP hook is invoked early in the receive path, before the packet traverses the full Linux networking stack. The XDP program reads current policy state from eBPF maps, performs concise packet parsing and policy checks, updates counters where needed, and applies an action. This path is intended to be deterministic and compact. It should not depend on synchronous user-space inference for each packet.

Candidate actions include XDP_PASS, XDP_DROP, XDP_REDIRECT where supported, traffic accounting, temporary blocklist lookup, source or device policy lookup, implementation-specific bounded rate control, and action-expiration checks. Packet dropping can be performed through XDP. Redirection may depend on driver, deployment, and network support. Rate control may use token-bucket-style counters in XDP or another appropriate Linux traffic-control mechanism. Quarantine may require routing, firewall, VLAN, redirect, SDN, or network-access-control integration. Administrator escalation is handled in user space rather than inside the XDP program. These distinctions are consistent with eBPF/XDP being a fast enforcement substrate rather than a complete orchestration system (Gregg, 2019; Høiland-Jørgensen et al., 2018; Vieira et al., 2020).

Table 6: Candidate Mitigation Actions and Enforcement Locations

| **Mitigation action** | **Primary enforcement location** | **eBPF/XDP role** | **Additional mechanism if required** | **Bounded-safety condition** |
|---|---|---|---|---|
| Pass | XDP data path | Apply XDP_PASS under current policy. | None in simple deployments. | Default action unless policy state indicates otherwise. |
| Monitor only | XDP and user space | Update counters or telemetry while passing packets. | Logs or dashboard may consume summaries. | No packet disruption; bounded telemetry volume. |

Table 6: Candidate Mitigation Actions and Enforcement Locations (continued)

| **Mitigation action** | **Primary enforcement location** | **eBPF/XDP role** | **Additional mechanism if required** | **Bounded-safety condition** |
|---|---|---|---|---|
| Temporary drop | XDP data path | Match temporary blocklist or policy state and apply XDP_DROP. | User-space controller sets and expires map state. | Time-limited; narrower scope preferred for critical devices. |
| Rate limit | XDP or Linux traffic control | May use counters or token-bucket-style logic. | tc, qdisc, firewall, or gateway shaping may be needed. | Rate and duration must be bounded and logged. |
| Destination-port restriction | XDP or firewall layer | Match selected ports and device policy in maps. | Firewall or routing policy may be used for complex cases. | Restrict only approved port ranges and expire automatically. |
| Redirect | XDP data path where supported | Apply XDP_REDIRECT or redirect-like handling. | Driver, routing, AF_XDP, or service-chain support may be required. | Deployment-dependent; must preserve rollback path. |
| Quarantine | External network controls | May mark or redirect selected traffic. | VLAN, NAC, SDN, firewall, or router integration may be required. | Requires high confidence or administrator approval for critical devices. |
| Administrator escalation | User-space control plane | No direct packet action required. | Alert dashboard, SIEM, email, or ticket workflow. | Used when uncertainty, criticality, or policy requires human review. |

## 7.3 Telemetry Collection and Resource-Aware Feature Extraction

The kernel path can maintain counters and export selected telemetry or flow summaries to user space through implementation-appropriate mechanisms such as eBPF maps, ring buffers, perf buffers, AF_XDP or another justified packet-sampling path, and flow aggregation in user space. The framework does not require full payload inspection by default. It focuses on packet-header-level and flow-level evidence to reduce privacy exposure and processing cost.

Candidate features include protocol, source and destination ports, packet rate, byte rate, flow duration, TCP flags, packet-size statistics, inter-arrival time, failed connection attempts, destination diversity, outbound connection bursts, and device-specific baseline deviations. The final feature set must be selected according to predictive value, extraction cost, latency, and live availability. Features that are useful in an offline dataset may be unsuitable if they require long windows, payload inspection, unavailable labels, or expensive state maintenance on a constrained gateway.

## 7.4 AI-Assisted Risk Scoring

The AI-assisted risk scorer estimates whether a flow, device, or short traffic window is likely to be benign, suspicious, or malicious. Binary risk scoring can support mitigation decisions, while multiclass classification can support reporting and incident analysis. The model generates a risk signal rather than a direct blocking command. The mitigation controller decides whether and how that signal should affect enforcement.

Candidate model classes include logistic regression, decision tree, random forest, gradient-boosted trees, autoencoder, and compact neural network. No single model is selected or validated in this conceptual manuscript. Confidence calibration matters because controller policy may require different responses for high-confidence attack evidence, ambiguous anomalies, and low-confidence deviations. Deployment cost also matters: feature-window design, incomplete feature windows, memory use, and inference latency affect responsiveness. A lower-complexity model may be preferable when it provides acceptable detection quality with substantially lower inference cost and more stable operational behaviour.

## 7.5 Policy-Mediated Mitigation Controller

The mitigation controller is the governance point between AI-assisted risk scoring and packet-level enforcement. It receives the risk score, confidence, device identifier, device role, device criticality, recent alert history, local allowlists, policy rules, current action state, expiration time, and administrator override status. It uses these inputs to select a bounded response from pre-approved action templates rather than allowing the model to generate arbitrary kernel logic.

Permitted responses include forwarding normally, monitoring only, increasing telemetry, generating an alert, applying a temporary rate limit, applying a temporary block, restricting selected destination ports, redirecting traffic where supported, requesting quarantine through external controls, escalating to an administrator, or rolling back an expired or disproven action. Disruptive actions must be time-limited, logged, and explicitly reversible. Broader actions require higher confidence or human confirmation, especially for healthcare, industrial, access-control, or other critical devices. Policy should fail conservatively when uncertainty is high: for example, by monitoring, alerting, or applying a narrow rate limit rather than broad blocking.

| State | Trigger | Bounded action | Exit or timeout | Critical-device rule | Required log evidence |
|---|---|---|---|---|---|
| Normal | No current risk condition | Forward normally; maintain counters | Risk threshold, policy change, or administrator request | Preserve service continuity | Device ID; policy version; baseline status |
| Observe | Low-confidence anomaly or drift signal | Increase telemetry; continue forwarding | Observation window or escalation threshold | Prefer observation over disruption | Risk score; confidence; observed features |
| Alert | Policy threshold met without immediate disruption | Generate alert; narrow monitoring if needed | Review, timeout, or stronger evidence | Escalate before disruption where required | Alert reason; device role; evidence summary |
| Temporarily rate-limited | Moderate or high risk with policy support | Apply bounded rate limit or request shaping | Expiry, override, or evidence update | Use narrower limits and shorter duration | Rate; duration; flow scope; map-update status |
| Temporarily blocked | High-confidence suspicious behaviour in non-critical scope | Apply temporary drop or destination restriction | Expiry, rollback, or administrator override | Require higher confidence or human confirmation | Block scope; timeout; confidence; rollback plan |
| Quarantine requested | Strong compromise indication or repeated severe events | Request external quarantine mechanism | Administrator decision or external-control status | Human confirmation preferred or required | Quarantine reason; external request; device criticality |
| Administrator escalation | Uncertainty, criticality, repeated alerts, or policy requirement | Notify administrator; preserve evidence | Administrator decision or policy timeout | Default for disruptive critical-device decisions | Full event record; explanation if available; recommended action |
| Rolled back | Timeout, override, disproven risk, policy change, or protected-device rule | Remove or narrow enforcement state | Map update and verification complete | Restore service unless policy prevents it | Rollback trigger; previous action; new state; verification status |

Table 7: Policy-Mediated Controller States and Bounded Transitions

These states are conceptual controller states for future implementation and validation. They define the intended governance semantics of the architecture rather than reporting a completed controller implementation.

## 7.6 Event-Level Explainability Without Blocking the Critical Path

Explanation generation should not necessarily delay urgent bounded mitigation. The controller can apply a cautious, reversible provisional action when risk and policy justify it. Explanation generation then proceeds asynchronously or in parallel, and the explanation is attached to the event log and administrator alert. The explanation can support confirmation, narrowing, extension, escalation, or rollback of the action.

Suspicious events and selected benign samples should receive explanations, but benign traffic does not require continuous expensive explanation. SHAP or LIME may be used depending on the selected model, but this manuscript does not imply that either method has already been implemented (Lundberg & Lee, 2017; Ribeiro et al., 2016). Explanations must be translated into operational narratives. For example: "A camera was temporarily rate-limited because its outbound SYN rate increased sharply, its destination diversity exceeded its baseline, and repeated connection attempts were observed within a short flow window." Explanation latency must be measured separately from detection-to-enforcement latency.

## 7.7 Logging, Rollback, and Model-Update Validation

Each event log should include timestamp, device identifier, flow identifier, relevant metadata, risk score, confidence, selected policy, explanation summary when available, action, action start time, expiration time, rollback status, administrator override, model version, and reason for escalation. Logs should support local review and future export to dashboards or SIEM workflows without requiring raw payload collection by default.

Rollback can occur automatically after timeout, after administrator override, after updated evidence, after explanation review, after policy change, or when a protected-device rule is triggered. Rollback should remove or narrow the relevant state in the eBPF maps or the external enforcement request and record the resulting state. Future model updates must be versioned, validated before enforcement, tested locally, and reversible. Federated-learning mechanisms may be used in future work, but updates must not directly alter kernel enforcement without policy validation (Khraisat et al., 2024; McMahan et al., 2017; Zhang et al., 2025).

## 7.8 Timing Semantics and Evaluation Boundaries

The architecture distinguishes several timing intervals: telemetry-aggregation latency, feature-extraction latency, inference latency, controller-decision latency, eBPF-map-update latency, detection-to-enforcement latency, explanation latency, and rollback latency. Detection-to-enforcement latency is defined as the interval between the time at which a user-space risk decision satisfies the controller's mitigation condition and the time at which the corresponding bounded enforcement state is successfully written to the relevant eBPF maps or applied through an external enforcement mechanism.

This metric does not imply blocking of the first malicious packet. Figure 2 illustrates the control workflow, including the immediate kernel fast path, user-space analysis path, and controlled response path. Explanation latency should be measured separately because explanations may be asynchronous and should not necessarily block urgent bounded action. Packet-processing overhead should be measured independently on the XDP fast path, including pass-through, accounting, map lookup, drop, redirect, and bounded rate-control cases where implemented. Timeouts and rollback timing also require separate validation because safe mitigation depends not only on how quickly an action is applied, but also on whether it expires, narrows, or reverses correctly.

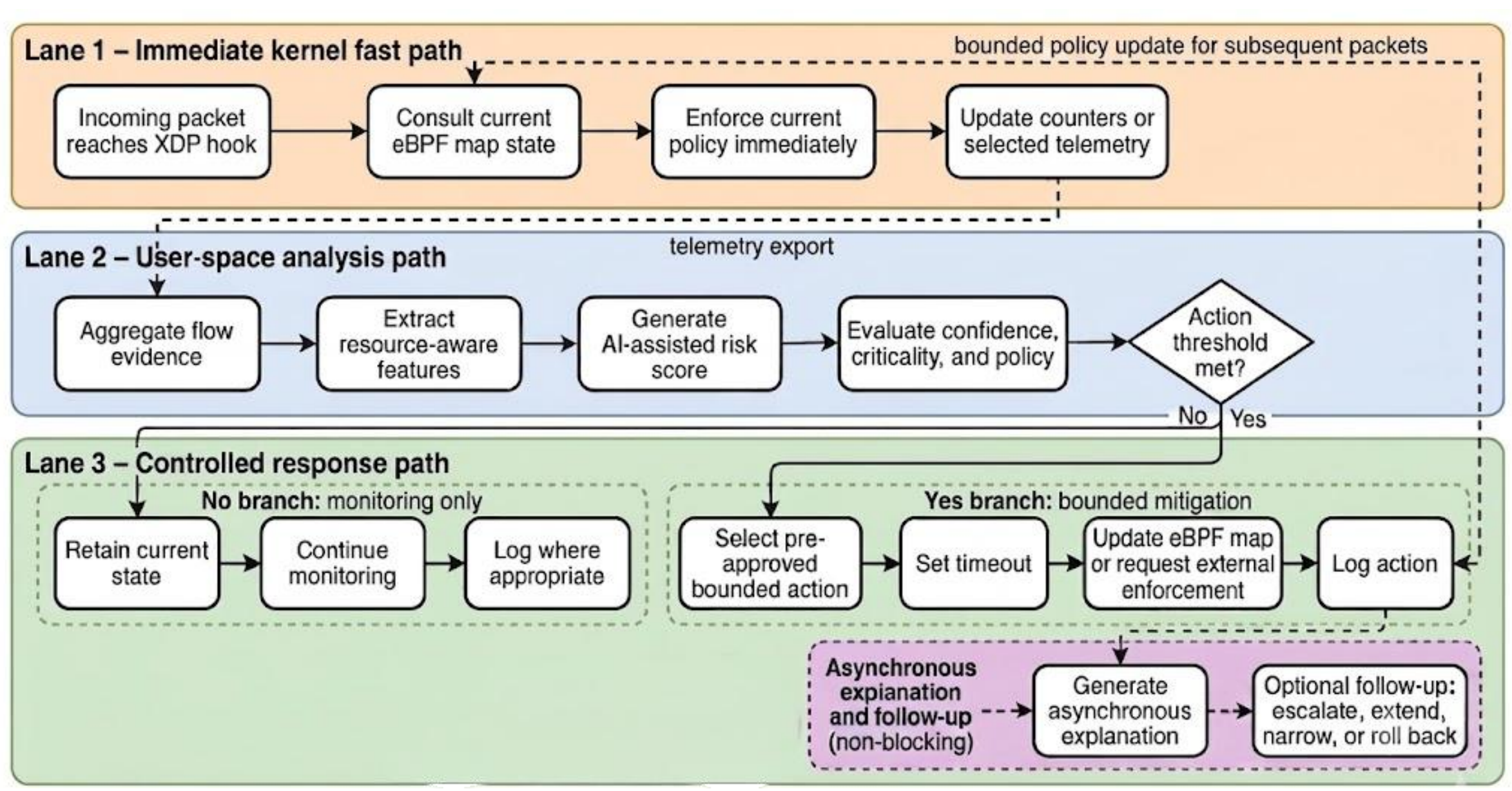

Figure 2: Control workflow showing immediate XDP enforcement of current policy state and asynchronous user-space analysis for bounded updates affecting subsequent packets.

# 8 Future Methodology and Reproducibility Plan

This section defines a compact protocol for future empirical validation of the proposed framework. It covers dataset experiments, Raspberry Pi-class deployment, bounded enforcement, XAI, safety, robustness, and reproducibility. No empirical results are claimed in this article.

## 8.1 Evaluation Questions

The protocol is organized around seven questions that connect offline detection evidence with deployable edge-system behaviour.

**RQ1. Detection quality**
How accurately can resource-aware models distinguish benign and malicious IoT behaviour across random, temporal, device-separated, attack-session-separated, and cross-dataset evaluations?

**RQ2. Deployment feasibility**
What telemetry, feature-extraction, inference, memory, and packet-processing costs arise on a Raspberry Pi-class gateway?

**RQ3. Enforcement responsiveness**
How quickly can a satisfied controller condition produce a verified bounded update in one or more eBPF maps or external enforcement mechanism?

**RQ4. Safety**
Do timeouts, rollback, critical-device policies, and conservative fallback behaviours preserve legitimate traffic?

**RQ5. Explainability**
Can event-level explanations support review without blocking urgent bounded mitigation or exhausting the gateway under alert bursts?

**RQ6. Robustness**
How does the framework behave under concept drift, low-rate evasion, spoofing indicators, telemetry flooding, map-capacity pressure, alert flooding, and poisoned update scenarios?

**RQ7. Reproducibility**
Can the complete protocol be repeated using documented code, configuration, topology, policy templates, and dataset-processing scripts?

## 8.2 Dataset Selection and Scope

Dataset evaluation should use CICIoT2023, TON_IoT, Edge-IIoTset, N-BaIoT, and BoT-IoT as IoT or IoT-adjacent datasets, with UNSW-NB15 only as a supplementary non-IoT-specific benchmark (Alsaedi et al., 2020; Koroniotis et al., 2019; Meidan et al., 2018; Moulahi et al., 2022; Moustafa & Slay, 2015; Neto et al., 2023). Table 8 summarizes intended roles without asserting unverified dataset attributes.

| Dataset | Scope | Intended evaluation role | Split considerations | Cross-dataset use | Main caveat |
| --- | --- | --- | --- | --- | --- |
| CICIoT2023 | IoT-focused | Primary dataset for broad IoT attack coverage and initial model comparison. | Use random stratified, temporal, device-separated, or attack-session-separated evaluation only where the official dataset structure supports the relevant split. | Use only after defensible feature alignment. | Dataset environment may not fully represent local deployment traffic. |
| TON_IoT | IoT and IIoT | Primary or complementary dataset for heterogeneous IoT and IIoT evaluation. | Confirm timestamps, device identifiers, and session structure from official documentation before selecting split regimes. | Use only after defensible feature alignment. | Heterogeneous feature definitions may restrict alignment. |
| Edge-IIoTset | IoT and IIoT | Complementary dataset for edge-oriented attack evaluation. | Use only split regimes supported by verified metadata and capture structure. | Use only after defensible feature alignment. | Testbed-specific traffic may limit generalization. |
| N-BaIoT | IoT botnet traffic | Complementary dataset for device-specific botnet detection analysis. | Confirm device and session structure before device-separated or session-separated evaluation. | Use only comparable live-available features. | Device and attack coverage may be narrower than more recent datasets. |
| BoT-IoT | Botnet-oriented IoT-adjacent traffic | Supplementary robustness benchmark. | Use split regimes supported by verified metadata. | Use only after defensible feature alignment. | Synthetic or testbed structure may affect deployability claims. |
| UNSW-NB15 | Non-IoT supplementary benchmark | Supplementary robustness comparison only. | Do not treat as primary IoT evidence. | Limited supplementary comparison. | Not IoT-specific. |

Table 8: Planned Dataset Roles and Evaluation Constraints

Before implementation, the exact availability of timestamps, device identifiers, session identifiers, and alignable features must be confirmed from the official dataset documentation. The evaluation protocol should report unavailable split regimes explicitly rather than forcing invalid comparisons.

## 8.3 Leakage Prevention and Data Splitting

Future evaluation must remove exact duplicates before splitting; identify near-duplicate or repeated flow records where feasible; prevent identical or closely related flows from appearing in both training and test partitions; fit scalers, encoders, feature selectors, dimensionality-reduction steps, and calibration models only on training data; apply class weighting or resampling only to training data; preserve the unseen test set until final evaluation; document preprocessing steps; report split sizes and class distributions; record random seeds; and avoid forced comparisons where dataset structure does not support a valid split.

Where feasible, the protocol should include random stratified, time-separated, device-separated, attack-session-separated, cross-dataset, and live-gateway evaluation. Random record-level splitting alone may overstate deployability because related flows, attack-session artefacts, or device-specific patterns can leak across partitions.

## 8.4 Feature Definition, Live Availability, and Window Design

A feature-definition file should record feature name, description, source, header-level or flow-level status, extraction location, aggregation window, state requirement, privacy implication, expected cost, live availability, dataset availability, and cross-dataset use. Candidate features include protocol, ports, TCP flags, packet and byte counts, packet and byte rates, flow duration, packet-size statistics, inter-arrival timing, failed-connection indicators where observable, destination diversity, outbound bursts, and device-specific baseline deviations.

Features requiring payload semantics, unavailable labels, or expensive long-term state should be excluded unless explicitly justified. Multiple flow-window sizes should be tested because shorter windows may improve responsiveness while longer windows may improve feature stability.

## 8.5 Model Families, Tuning, and Calibration

Candidate models should include logistic regression, decision tree, random forest, gradient-boosted trees, autoencoder, and compact neural network. Selection should be Pareto-efficient rather than based only on maximum F1-score, balancing detection quality with feature availability, inference cost, memory footprint, explanation cost, and calibration.

Reports should document model type, hyperparameters, training time, model size, inference runtime, memory footprint, feature set, calibration method, confidence threshold, seed, and software versions. Metrics should include macro-F1, weighted F1, per-class recall, false-positive rate, false-negative rate, confusion matrix, precision-recall curves where useful, calibration curve, and Brier score or expected calibration error where feasible. Calibration is necessary because controller actions vary by risk and uncertainty.

## 8.6 Hardware Testbed and Deployment Topology

The testbed report should specify Raspberry Pi model, RAM, storage, operating system, kernel version, NIC model, topology, routing or bridge mode, VLAN use, IPv4 and IPv6 behaviour, wired or wireless path, XDP mode, NIC offload settings, compiler, BCC/libbpf/bpftool or equivalent tooling, eBPF map types and capacities, telemetry export, user-space runtime, model runtime, flow-window configuration, traffic-generator host, monitoring host, container topology, and physical IoT devices where available.

The topology must state which traffic crosses the gateway, whether east-west traffic is observable, what bypasses enforcement, whether the interface is a bottleneck, whether redirect is supported, and whether quarantine uses XDP, Linux traffic control, firewall, routing, VLAN, network-access control, software-defined networking, or another supported mechanism.

## 8.7 Traffic Workloads and Attack Scenarios

Benign workloads should include iperf3, web requests, MQTT-style telemetry, DNS, periodic device communication, bursty legitimate traffic, device background traffic, and mixed concurrent flows. Controlled malicious scenarios should include scanning, brute-force behaviour where observable, UDP flood, TCP SYN flood, low-rate DoS, spoofing indicators, botnet-like outbound traffic, unusual destination diversity, lateral movement across observable paths, telemetry flooding, alert flooding, map-capacity pressure, explanation-workload pressure, poisoned update scenarios, and benign drift. All tests must remain in authorized laboratory environments.

## 8.8 Baseline Matrix

Table 9 defines the baseline matrix. Rollback-disabled and safeguard-disabled configurations are laboratory-only comparisons, not operational recommendations.

| Baseline ID | Configuration | Purpose | Key measurements | Main comparison |
|---|---|---|---|---|
| B0 | Native gateway forwarding without XDP attachment | Ordinary forwarding cost | Throughput, loss, CPU, memory, latency | All XDP-attached configurations |
| B1 | XDP pass-through only | Attachment and pass-path overhead | Packet-processing overhead, forwarding continuity | B0 and telemetry configurations |
| B2 | XDP accounting and telemetry only | Observation overhead without mitigation | Counter accuracy, telemetry drops, CPU, map occupancy | B1 and controller-enabled designs |
| B3 | Threshold-only detection with bounded enforcement | Simple policy-driven mitigation | Detection events, false blocking, enforcement latency, rollback | AI-assisted scoring |
| B4 | AI-assisted scoring without enforcement | Detection and inference costs | Detection metrics, calibration, inference latency, memory | B3 and controller-enabled variants |
| B5 | AI-assisted scoring plus bounded controller without XAI | Governed mitigation without explanation overhead | Controller latency, map-update latency, safety outcomes | B6 and B8 |
| B6 | AI-assisted scoring plus controller plus asynchronous XAI | Explanation support outside urgent mitigation path | Explanation latency, queueing, controller timing, resource use | B5 |
| B7 | Complete design with rollback disabled for controlled safety comparison | Laboratory-only rollback contrast | False blocking duration, recovery behaviour, service impact | Rollback-enabled design |
| B8 | Complete design with critical-device policies enabled | Policy differentiation by device class | Escalation, narrower mitigation, traffic preservation | Standard policy treatment |
| B9 | Complete design under resource-exhaustion stress | Bounded behaviour during overload | Drops, queue pressure, map occupancy, fallback, recovery | Non-stressed complete design |

Table 9: Future Experimental Baselines

## 8.9 Ablation Studies

Ablations should change one component at a time where feasible and should measure effects without predicting results.

| Ablation | Changed component | Purpose | Measurements | Interpretation boundary |
|---|---|---|---|---|
| Feature-set size | Number and type of features | Cost and detection trade-off | Detection, extraction latency, memory | Live-available features only |
| Flow-window duration | Aggregation window | Responsiveness versus stability | Detection delay, feature variance, false positives | Workload-dependent |
| Model complexity | Model family or size | Cost and accuracy comparison | Inference latency, model size, calibration, F1 | Accuracy is not sufficient alone |
| Telemetry-export mechanism | Export path or buffer type | Observation overhead | Drops, latency, CPU, queue pressure | Tooling-specific |
| XAI disabled | Explanation module | Explanation overhead isolation | Resource use, review support, event timing | Does not test usefulness |
| Asynchronous versus synchronous XAI | Explanation scheduling | Mitigation-path blocking risk | Controller latency, explanation latency, queues | Synchronous mode may be laboratory-only |
| Rollback disabled | Rollback logic | Controlled safety contrast | Recovery, false blocking duration, service impact | Laboratory-only |
| Critical-device policies disabled | Device policy layer | Safeguard contribution | Traffic preservation, escalation | Laboratory-only for sensitive classes |
| Drift checks disabled | Drift-monitoring layer | Drift-detection value | Temporal performance, fallback | Scenario-dependent |
| Alert suppression disabled | Alert-rate control | Attention and queue overload | Alert volume, delayed alerts, CPU, storage | Human conclusions require user study |
| Map-capacity bounds varied | eBPF map capacity | Bounded state management | Occupancy, insertion failure, fallback, recovery | Hardware-specific |
| Threshold-only versus AI-assisted scoring | Detection logic | Simple versus adaptive risk signals | Detection, calibration, enforcement actions | Equivalent features where possible |
| XDP native versus generic mode | XDP attachment mode | Mode-dependent overhead | Throughput, packet rate, CPU, drops | Only where supported |
| Redirect or rate-control mechanism varied | Enforcement backend | Quarantine and throttling options | Enforcement latency, loss, recovery | Deployment-supported mechanisms only |

Table 10: Future Ablation Studies

## 8.10 Timing and Systems Measurements

Timing should separate telemetry aggregation, feature extraction, inference, controller decision, eBPF-map-update latency, detection-to-enforcement, explanation, and rollback. Detection-to-enforcement latency is the interval between the time at which a user-space risk decision satisfies the controller's mitigation condition and the time at which the corresponding bounded enforcement state is successfully written to the relevant eBPF maps or applied through an external enforcement mechanism.

Explanation latency should be reported separately, first-packet blocking is not assumed, and packet-processing overhead should be measured independently. Reports should include median, p95, and p99 timing values, plus throughput, packet rate, packet loss, CPU, memory, queue pressure, and map occupancy.

## 8.11 Safety and Critical-Device Evaluation

Safety testing should cover timeout enforcement, rollback success and verification, administrator override, allowlists, policy-template enforcement, conservative fallback, false blocking, legitimate-traffic preservation, critical-device escalation, narrower mitigation, and service recovery. Results should be reported for standard, sensitive, critical, and safety-critical policy classes. Safety-critical disruptive actions should be simulated or laboratory-only.

## 8.12 Resource-Exhaustion Evaluation

Stress tests should examine telemetry buffer pressure, ring-buffer or perf-buffer drops where used, eBPF-map occupancy, map-capacity exhaustion, log volume, storage pressure, alert flooding, explanation queueing, CPU pressure, memory pressure, administrator-attention overload, and recovery. Reports should include bounded-capacity settings, dropped telemetry, delayed alerts, policy-state integrity, fallback behaviour, forwarding continuity, and recovery time.

## 8.13 Concept Drift and Adversarial Robustness

Robustness evaluation should separate benign drift, inference-time evasion, low-rate evasion, feature manipulation, source spoofing indicators, poisoned model updates, non-IID federated updates, and malicious-client influence. Tests should include temporal windows, changed destinations, changed schedules, firmware-like behaviour changes, slow-rate attacks, update acceptance and rejection, rollback to a previous model, and conservative fallback. Complete adversarial robustness should not be claimed without evidence.

## 8.14 XAI Evaluation

XAI evaluation should report explanation latency, queueing delay, stability under similar events, perturbation stability, feature relevance, faithfulness where feasible, sparsity, administrator usefulness, agreement with known domain indicators, false-confidence risk, and asynchronous execution status. SHAP, LIME, or model-specific methods may be compared depending on the selected model (Lundberg & Lee, 2017; Ribeiro et al., 2016). Explanations should support review without delaying urgent bounded mitigation unless synchronous use is explicitly justified.

## 8.15 Statistical Reporting

Future studies should use repeated runs, fixed seeds, documented repetition counts, median, mean where useful, standard deviation or interquartile range, p95, p99, confidence intervals where feasible, and effect sizes for comparative claims. Dataset and live-gateway experiments should be reported separately. Best-case results alone are insufficient.

## 8.16 Reproducibility Artefacts

Future implementation should release or archive source code, eBPF/XDP programs, user-space controller logic, policy templates, preprocessing scripts, feature definitions, dependency versions, testbed topology, workload commands, seeds, processed results, and an ethical-use statement. Appendix A provides the complete reproducibility checklist.

## 8.17 Evaluation-Metrics Summary

Table 11 summarizes the required metric categories using compact reporting targets.

| Category | Metrics | Source | Notes |
|---|---|---|---|
| Detection | Macro-F1; weighted F1; per-class recall; FPR; FNR; confusion matrix; PR curves | Dataset splits; labelled live scenarios | By split and attack class |
| Calibration | Calibration curve; Brier score; ECE where feasible; threshold behaviour | Validation and test data | Separate from accuracy |
| Deployment | Extraction latency; inference latency; model size; memory; CPU | Runtime instrumentation; system monitors | Hardware details required |
| Enforcement | Controller latency; map-update latency; detection-to-enforcement; fast-path overhead; throughput; loss | Controller logs; XDP counters; traffic generator | Explanation reported separately |
| Safety | Timeout; rollback; false blocking; traffic preservation; recovery; override | Controller logs; probes; policy audit | By policy class |
| XAI | Latency; queueing; stability; relevance; faithfulness; sparsity; usefulness | XAI logs; perturbation tests; review instruments | State async or sync |
| Robustness | Drift response; evasion; spoofing indicators; update rejection; model rollback; fallback | Drift, adversarial, and update tests | No complete-robustness claim |
| Resource exhaustion | Buffer drops; map occupancy; log volume; alert delay; queue length; recovery time | eBPF map; buffers; logs; monitors | Include capacity settings |
| Reproducibility | Code; configuration; topology; seeds; versions; results; plots | Artefact checklist | Missing items explicit |

Table 11: Future Evaluation-Metrics Summary

# 9 Discussion

## 9.1 Adaptability versus Operational Safety

Threshold rules remain useful for clear traffic anomalies such as floods, repeated connection attempts, or obvious volume spikes. They are transparent and inexpensive, but static rules can miss low-rate behaviour and multivariate deviations. AI-assisted scoring can capture combinations of flow features that are difficult to express as single thresholds, yet it also introduces uncertainty, calibration needs, and false-positive risk. The framework therefore treats AI output as a risk signal rather than an autonomous enforcement authority. Policy mediation, bounded actions, timeouts, rollback, critical-device rules, and escalation are intended to make adaptive mitigation auditable and reversible rather than unconstrained.

## 9.2 Fast Enforcement versus Rich Reasoning

XDP supports early packet handling and is suitable for concise actions such as pass, drop, rate-control support, accounting, or redirection once bounded state has been established. Complex reasoning belongs in user space, where model scoring, policy arbitration, explanation scheduling, logging, and rollback can be inspected and governed. This two-plane separation supports maintainability while preserving a fast enforcement path. It also avoids assuming first-packet blocking: detection-to-enforcement latency and packet-processing overhead remain distinct measurements.

## 9.3 Explainability versus Resource Cost

Explanations can support triage, rollback review, administrator trust, and policy refinement. They can also consume CPU, memory, queue capacity, and attention if generated for every flow or placed in the urgent mitigation path. The intended compromise is asynchronous event-level explanation: bounded mitigation may proceed according to policy and calibrated risk, while explanations support review and escalation. Future evaluation must therefore test usefulness, latency, queueing, stability, agreement with known indicators, and false-confidence risk.

### 9.4 Adaptation versus Governance

Concept drift, local retraining, federated updates, and TinyML extensions may improve adaptability as devices, firmware, cloud dependencies, and attacks change. They also create poisoning, non-IID data, versioning, rollback, acceptance-testing, and validation risks. Model updates must not bypass policy checks or critical-device safeguards, and TinyML or federated learning should remain extensions until their security and resource costs are measured. Overall, the contribution is a governed coordination model: adaptive detection, explanation, bounded policy, rollback, and packet-level enforcement must work together without overloading constrained gateways or disrupting legitimate traffic.

## 10 Practical Implications

### 10.1 Consumer and Small-Organization Gateways

Homes, schools, and small organizations often operate cameras, access-control devices, sensors, appliances, and classroom equipment with limited security staff. A low-cost Linux gateway could add local monitoring, understandable logs, and bounded actions as a complement to router configuration, segmentation, and firewall rules. This use case requires validation under ordinary benign bursts, cloud-dependent device behaviour, and mixed user routines.

### 10.2 Institutional and Safety-Sensitive Environments

Healthcare, physical-access, and safety-related environments require stricter device classes because false blocking can affect clinical workflows, entry control, or operational continuity. Sensitive and critical devices may need narrower mitigation, faster rollback, allowlist handling, or human confirmation. In these settings, alerting or rate limiting may be safer than immediate blocking for some devices, and deployment would require local policy review.

### 10.3 Industrial and Distributed Infrastructure

Industrial IoT and smart-city deployments may benefit from localized enforcement near production networks or distributed sensing sites, reducing dependence on continuous cloud connectivity and supporting site-specific policy. These environments vary in topology, protocol use, maintenance windows, and availability requirements. Privacy-aware model updates could support multi-site learning, but only after validation of security, governance, and operational cost.

# 11 Limitations

## 11.1 Conceptual and Empirical Limitations

This article reports a conceptual framework and research agenda, not a prototype or deployed system. It claims no latency, throughput, detection, rollback, explanation, resource-consumption, or mitigation measurements. Architecture alone cannot prove deployability; it only defines what future implementation should test.

## 11.2 Dataset and Generalization Limitations

Public datasets may contain biases, synthetic artefacts, capture assumptions, mismatched labels, feature inconsistencies, imbalance, and repeated flows. Random record-level splitting may overstate performance when related flows or attack sessions cross partitions. Temporal, device-separated, session-separated, and cross-dataset tests reduce this risk only when metadata supports them, and none can replace live-gateway testing.

## 11.3 Deployment and Visibility Limitations

The framework observes only traffic crossing the enforcement point. Bypass paths, unobserved east-west traffic, wireless or switched segments, and encrypted payloads can limit visibility. Hardware, drivers, topology, IPv4/IPv6 behaviour, bridge or routing mode, VLANs, multi-interface paths, and XDP mode can affect performance. Raspberry Pi-class results would not represent all gateways or NIC capabilities.

## 11.4 Adversarial and Operational Limitations

Low-rate attacks, evasion, spoofing indicators, map pressure, telemetry floods, logging floods, alert floods, XAI workload, and administrator-attention overload may reduce effectiveness. The framework also assumes the kernel, root administrator, and policy controller are not fully compromised. Bounded actions, allowlists, timeouts, and override reduce but do not eliminate false-positive risk.

### 11.5 Governance and Update Limitations

Drift detection, local retraining, federated learning, model versioning, rollback, and acceptance testing require validation before updates influence enforcement. Explanations may be technically correct but operationally unhelpful or unstable. Privacy obligations vary by deployment and jurisdiction, so the framework is not a legal-compliance claim.

## 12 Future Work

Future work should follow Section 8 through six priorities: implement a Raspberry Pi-class prototype with a kernel-resident XDP data path and user-space control plane; validate it under mixed benign and malicious traffic; evaluate larger topologies covering VLAN, bridge, routing, IPv6, wireless, multi-interface, and east-west visibility scenarios; stress rollback, critical-device policies, map capacity, telemetry pressure, alert flooding, explanation queues, and recovery; test drift, evasion, poisoning, privacy-aware updates, and TinyML extensions without bypassing policy validation; and integrate structured logs and alerts with dashboards or SIEM workflows.

## 13 Conclusion

This article is a conceptual framework and research agenda for explainable AI-assisted and policy-bounded eBPF/XDP mitigation at the IoT edge. Its contribution is architectural and operational rather than algorithmic: it defines a cautious coordination model, not a new classifier, XDP primitive, or completed implementation.

The framework separates a kernel-resident XDP data path from a user-space control and intelligence plane. The data path applies concise packet actions from current enforcement state, while user space handles telemetry aggregation, AI-assisted scoring, policy mediation, explanation scheduling, logging, rollback, and model-update control. AI output is treated as a risk signal rather than an autonomous blocking command.

Policy mediation, timeouts, rollback, structured evidence, administrator escalation, explanation, and critical-device safeguards constrain mitigation, but they remain design commitments that require implementation and hardware-aware validation. No measured superiority, proven real-time response, validated mitigation effectiveness, or completed empirical deployment is claimed.

# Appendix A: Reproducibility Checklist for Future Implementation

Table 12 lists the minimum artefacts for independent repetition, subject to ethical, legal, and dataset-licensing constraints.

Table 12: Reproducibility Artefacts for Future Implementation

| Artefact | Required content | Purpose | Availability statement |
|---|---|---|---|
| Source code | Repository; commits; build instructions | Audit and repetition | Release or controlled access |
| eBPF/XDP program | Packet logic; build flags | Data-path reproduction | Source and verifier-relevant settings |
| User-space controller | Risk handling; policy mediation; logging | Mitigation workflow | Runtime and dependencies |
| Policy templates | Standard; sensitive; critical; safety-critical | Bounded actions | Safe templates without secrets |
| Rollback logic | Timeouts; verification procedures | Safety checks | Test commands and log fields |
| Model-training scripts | Training; tuning; calibration; evaluation | Dataset experiments | Seeds and software versions |
| Preprocessing scripts | Cleaning; leakage checks; splitting; encoding | Dataset partitions | Split manifests where permitted |
| Feature-definition file | Names; windows; sources; costs | Offline/live alignment | Versioned with results |
| Configuration files | Runtime; model; policy; logging; topology | Deployment state | Credentials removed |
| Docker files | Images; service definitions | Client/service reproduction | Image digests where feasible |

*Continued on next page*

Table 12: Reproducibility Artefacts for Future Implementation (continued)

| Artefact | Required content | Purpose | Availability statement |
|---|---|---|---|
| Dependency versions | Libraries; compiler; tools; model runtime | Software control | Lock files or manifests |
| Kernel version | Kernel; headers; XDP mode; NIC driver | eBPF/XDP behaviour | Exact versions and settings |
| Map definitions | Types; keys; values; capacities | Enforcement state | Capacity and eviction assumptions |
| Testbed topology | Physical/logical diagrams | Traffic visibility | Bypass and east-west limits |
| Traffic-generation commands | Benign and malicious workloads | Live-test repetition | Authorized laboratory use only |
| Experiment commands | End-to-end commands; parameters | Measurement repetition | Timestamps and run IDs |
| Random seeds | Dataset; model; workload seeds | Repeatability | All fixed seeds reported |
| Processed results | Aggregated metrics; logs | Verification | Share within licence limits |
| Plotting scripts | Figures; tables; summaries | Reporting reproduction | Versioned with results |
| Data-availability statement | Dataset sources; access limits | Reuse constraints | Original datasets and licences |
| Limitations statement | Validity and topology limits | Prevent overclaiming | Published with results |
| Ethical-use statement | Authorization; containment | Misuse reduction | Laboratory-only attack testing |